\documentclass[RNAAS]{aastex631}

\begin{document}

\title{MPS-ATLAS library of stellar model atmospheres and spectra}

\author{N.~Kostogryz}
\affiliation{Max-Planck-Institut f\"ur Sonnensystemforschung, Justus-von-Liebig-Weg 3, 37077 G\"ottingen, Germany\\ }

\author{A.~I~Shapiro}
\affiliation{Max-Planck-Institut f\"ur Sonnensystemforschung, Justus-von-Liebig-Weg 3, 37077 G\"ottingen, Germany\\ }

\author{V.~Witzke}
\affiliation{Max-Planck-Institut f\"ur Sonnensystemforschung, Justus-von-Liebig-Weg 3, 37077 G\"ottingen, Germany\\ }

\author{D.~Grant}
\affiliation{School of Physics, University of Bristol, HH Wils Physics Laboratory, Tyndall Avenue, Bristol BS8 1TL, UK}

\author{H.~R.~Wakeford}
\affiliation{School of Physics, University of Bristol, HH Wils Physics Laboratory, Tyndall Avenue, Bristol BS8 1TL, UK}

\author{K.~B.~Stevenson}
\affiliation{Johns Hopkins APL, 11100 Johns Hopkins Road, Laurel, MD 20723, USA}

\author{S.~K.~Solanki}
\affiliation{Max-Planck-Institut f\"ur Sonnensystemforschung, Justus-von-Liebig-Weg 3, 37077 G\"ottingen, Germany\\ }

\author{L.~Gizon}
\affiliation{Max-Planck-Institut f\"ur Sonnensystemforschung, Justus-von-Liebig-Weg 3, 37077 G\"ottingen, Germany\\ }
\affiliation{Institut f\"ur Astrophysik, Georg-August-Universit\"at G\"ottingen, Friedrich-Hund-Platz 1, 37077 G\"ottingen, Germany}

\begin{abstract}
Stellar spectra contain a large amount of information about the  conditions in stellar atmospheres. However, extracting this information is challenging and demands comprehensive numerical modelling. 
Here, we present stellar spectra synthesized using  the recently developed state-of-the-art MPS-ATLAS code on a fine grid of stellar fundamental parameters. 
These calculations have been extensively validated against solar and stellar observations and can be used for various astrophysical applications. The spectra are available at the Max Planck Digital Library (MPDL, https://edmond.mpdl.mpg.de/dataset.xhtml?persistentId=doi:10.17617/3.NJ56TR) and have been also incorporated into the 
 ExoTiC-LD python package (https://github.com/Exo-TiC/ExoTiC-LD/ which returns stellar limb darkening coefficients used for the software package Exoplanet Timeseries Characterisation (ExoTic).
\end{abstract}


\section{Introduction} \label{sec:intro}

 Libraries of stellar synthetic spectra and their center-to-limb variations are often needed in the analysis of observed spectroscopic and photometric data for various applications, e.g.~for determining stellar fundamental parameters and characterizing their orbiting planets. A number of spectral libraries have been developed, e.g.~ATLAS9 \citep{1993KurCD..13.....K}, NextGen \citep{1999ApJ...512..377H}, MARCS \citep{2008A&A...486..951G}, Phoenix: G\"ottingen Spectral Library \citep{2013A&A...553A...6H}, and STAGGER \citep{2015A&A...573A..90M}. In this paper, we present a new library of stellar specific intensity spectra at different disk positions as well as disk-integrated flux spectra computed with the MPS-ATLAS code \citep{2021A&A...653A..65W}. We use the same grid of effective temperatures  ($T_{\rm{eff}}$), metallicities ($M/H$), and surface gravities ($\log g$) as \cite{2022A&A...666A..60K} who recently used MPS-ATLAS to calculate stellar limb darkening coefficients. This grid is substantially finer than other published grids especially in metallicity values. 

\section{Atmospheric models and stellar spectra}\label{sec:models}

We have recently developed the one-dimensional (1D) Merged Parallelized Simplified ATLAS code \citep[MPS-ATLAS][]{2021A&A...653A..65W}.  It is based on the two codes ATLAS9 \citep{1993KurCD..13.....K} and DFSYNTHE \citep{2005MSAIS...8...34C, 2005MSAIS...8...86K} which we have greatly updated. MPS-ATLAS has been intensively tested against solar and stellar observations and incorporates improved molecular and atomic data \citep{2021A&A...653A..65W, 2022A&A...666A..60K}. It incorporates the implementation of the optimized opacity distribution functions \citep[ODF,][]{2019A&A...627A.157C, 2021ApJS..255....3A} as well as MPI parallelization, which allowed  us to perform calculations on a fine grid of stellar fundamental parameters. 

\cite{2022A&A...666A..60K} presented a new stellar limb darkening library which demonstrates excellent agreement with available solar \citep{1994SoPh..153...91N} and stellar data \cite{2018A&A...616A..39M, 2023MNRAS.519.3723M}. In \cite{2022A&A...666A..60K} we only provide limb darkening coefficients computed for passbands of broadband photometric missions, such as {\it Kepler}, TESS, CHEOPS, and PLATO\footnote{This work is not an official PLATO mission deliverable}. In this paper, we provide the stellar spectra at 24 positions from the disk center to the limb and disk-integrated spectra (obtained from spectra calculated at different disk positions) together with the corresponding atmospheric models. The spectra are calculated on 1221 spectral intervals taken from  \cite{1993KurCD..13.....K}. The calculations are performed on an extensive and fine grid of fundamental stellar parameters. Namely, the temperature grid  covers the range from 3500~K to 9000~K with a step of 100~K. We further consider the range of gravity values $3.0 \le \log g \le 5.0$, which corresponds to stars from dwarfs to sub-giants. The most extensive grid is on stellar metallicity and covers a range from $-5.0$ to $1.5$ with a step of $0.05$ for $-1\le M/H \le 0.5$, a step of $0.5$ for $ M/H \le -2.5$, and a step of $0.1$ for the remaining metallicity values. A detailed description of the grid is given in \cite{2022A&A...666A..60K}. 

We present two sets of model atmospheres and spectra, `set~1' is based on the  chemical element composition by \cite{1998SSRv...85..161G}  and `set~2' on the composition by \cite{2009ARAA..47..481A}. We used mixing-length parameter of $1.25$ for  `set~1' and implemented values from \cite{2018ApJ...858...28V} for `set~2'. We allowed a convective overshoot of one pressure scale height for most of the models. Line blanketing was taken into account by using the Opacity Distribution Function(ODF) approach \citep{1984mrt..book..395C,2019A&A...627A.157C}. For the ODF computations we considered more than 100 million atomic and molecular transitions from \cite{2005MSAIS...8...86K}. The shape of spectral lines is approximated by the Voigt profile and broadened using a micro-turbulent velocity of 2~km/s \citep[see][]{2005MSAIS...8...86K, 2021A&A...653A..65W}. We include the continuum opacity contributions from both absorption and scattering processes \citep[see details in][]{2021A&A...653A..65W, 2022A&A...666A..60K}.

Both sets of stellar model atmospheres and spectra are publicly available on the MPDL (https://edmond.mpdl.mpg.de/dataset.xhtml?persistentId=doi:10.17617/3.NJ56TR). In addition to the atmosphere structure (i.e.~column mass, pressure and temperature) the models contain important input parameters, such as stellar parameters, overshoot, mixing-length parameter, and abundances of all chemical elements. We also include the convergence criteria, which are given by the maximum temperature change over all depth points between the two last iterations of the atmospheric structure calculations. For the subsequently computed spectra, we use the same settings, namely the same ODF, continuum opacity, equation-of-state and abundances, as were used for computing the atmospheric models.

\section{Integration into ExoTiC-LD}
ExoTiC-LD \citep{david_grant_2022_7437681} is a package developed from ExoTiC-ISM \citep{Laginja2020, iva_laginja_2020_3923986}. It produces stellar limb-darkening coefficients for specific instruments, stars, and wavelength ranges based on ATLAS9 \citep{2000A&A...363.1081C} and STAGGER \citep{2015A&A...573A..90M} libraries. We have incorporated our library (both `set~1' and `set~2') into ExoTiC-LD allowing users to obtain limb darkening coefficients with reduced errors for many stars \citep[due to the fine grid of fundamental parameters we used and high accuracy of our calculations, see][]{2021A&A...653A..65W, 2022A&A...666A..60K, 2023MNRAS.519.3723M}. 
Recently, \cite{2023MNRAS.519.3723M} obtained stellar limb darkening coefficients from {\it Kepler} and TESS observations and compared them to the coefficients computed with different existing spectra libraries. He found that MPS-ATLAS `set~1' leads to a better agreement with the observed stellar limb darkening curves.
Therefore, in ExoTiC-LD v3.0.0 \citep{david_grant_2022_7437681} we set the MPS-ATLAS `set~1' to be the default limb darkening library. 

\begin{acknowledgments}
N.K. and L.G. acknowledge support from the German Aerospace Center (DLR FKZ~50OP1902). V.W. and A.I.S. were funded by the European Research Council (ERC) under the European Union’s Horizon 2020 research and innovation program (grant no. 715947). L.G. acknowledges funding from ERC Synergy Grant WHOLE~SUN~810218 and from NYUAD Institute Grant G1502. D.G. acknowledges funding from the UKRI STFC Consolidated Grant ST/V000454/1.
\end{acknowledgments}

\bibliography{kostogryz}{}

\begin{thebibliography}{}
\expandafter\ifx\csname natexlab\endcsname\relax\def\natexlab#1{#1}\fi
\providecommand{\url}[1]{\href{#1}{#1}}
\providecommand{\dodoi}[1]{doi:~\href{http://doi.org/#1}{\nolinkurl{#1}}}
\providecommand{\doeprint}[1]{\href{http://ascl.net/#1}{\nolinkurl{http://ascl.net/#1}}}
\providecommand{\doarXiv}[1]{\href{https://arxiv.org/abs/#1}{\nolinkurl{https://arxiv.org/abs/#1}}}

\bibitem[{{Anusha} {et~al.}(2021){Anusha}, {Shapiro}, {Witzke}, {Cernetic},
  {Solanki}, \& {Gizon}}]{2021ApJS..255....3A}
{Anusha}, L.~S., {Shapiro}, A.~I., {Witzke}, V., {et~al.} 2021, \apjs, 255, 3,
  \dodoi{10.3847/1538-4365/abfb72}

\bibitem[{{Asplund} {et~al.}(2009){Asplund}, {Grevesse}, {Sauval}, \&
  {Scott}}]{2009ARAA..47..481A}
{Asplund}, M., {Grevesse}, N., {Sauval}, A.~J., \& {Scott}, P. 2009, \araa, 47,
  481, \dodoi{10.1146/annurev.astro.46.060407.145222}

\bibitem[{{Carbon}(1984)}]{1984mrt..book..395C}
{Carbon}, D.~F. 1984, in Methods in Radiative Transfer, ed. W.~Kalkofen
  (Cambridge University Press), 395--426

\bibitem[{{Castelli}(2005)}]{2005MSAIS...8...34C}
{Castelli}, F. 2005, Memorie della Societa Astronomica Italiana Supplementi, 8,
  34

\bibitem[{{Cernetic} {et~al.}(2019){Cernetic}, {Shapiro}, {Witzke}, {Krivova},
  {Solanki}, \& {Tagirov}}]{2019A&A...627A.157C}
{Cernetic}, M., {Shapiro}, A.~I., {Witzke}, V., {et~al.} 2019, \aap, 627, A157,
  \dodoi{10.1051/0004-6361/201935723}

\bibitem[{{Claret}(2000)}]{2000A&A...363.1081C}
{Claret}, A. 2000, \aap, 363, 1081

\bibitem[{Grant \& Wakeford(2022)}]{david_grant_2022_7437681}
Grant, D., \& Wakeford, H.~R. 2022, Exo-TiC/ExoTiC-LD: ExoTiC-LD v3.0.0,
  v3.0.0,  Zenodo, \dodoi{10.5281/zenodo.7437681}

\bibitem[{{Grevesse} \& {Sauval}(1998)}]{1998SSRv...85..161G}
{Grevesse}, N., \& {Sauval}, A.~J. 1998, Space Sci. Rev., 85, 161,
  \dodoi{10.1023/A:1005161325181}

\bibitem[{{Gustafsson} {et~al.}(2008){Gustafsson}, {Edvardsson}, {Eriksson},
  {J{\o}rgensen}, {Nordlund}, \& {Plez}}]{2008A&A...486..951G}
{Gustafsson}, B., {Edvardsson}, B., {Eriksson}, K., {et~al.} 2008, \aap, 486,
  951, \dodoi{10.1051/0004-6361:200809724}

\bibitem[{{Hauschildt} {et~al.}(1999){Hauschildt}, {Allard}, \&
  {Baron}}]{1999ApJ...512..377H}
{Hauschildt}, P.~H., {Allard}, F., \& {Baron}, E. 1999, \apj, 512, 377,
  \dodoi{10.1086/306745}

\bibitem[{{Husser} {et~al.}(2013){Husser}, {Wende-von Berg}, {Dreizler},
  {Homeier}, {Reiners}, {Barman}, \& {Hauschildt}}]{2013A&A...553A...6H}
{Husser}, T.~O., {Wende-von Berg}, S., {Dreizler}, S., {et~al.} 2013, \aap,
  553, A6, \dodoi{10.1051/0004-6361/201219058}

\bibitem[{{Kostogryz} {et~al.}(2022){Kostogryz}, {Witzke}, {Shapiro},
  {Solanki}, {Maxted}, {Kurucz}, \& {Gizon}}]{2022A&A...666A..60K}
{Kostogryz}, N.~M., {Witzke}, V., {Shapiro}, A.~I., {et~al.} 2022, \aap, 666,
  A60, \dodoi{10.1051/0004-6361/202243722}

\bibitem[{{Kurucz}(1993)}]{1993KurCD..13.....K}
{Kurucz}, R. 1993, ATLAS9 Stellar Atmosphere Programs and 2 km/s grid. Kurucz
  CD-ROM No. 13. Cambridge, 13

\bibitem[{{Kurucz}(2005)}]{2005MSAIS...8...86K}
{Kurucz}, R.~L. 2005, Memorie della Societa Astronomica Italiana Supplementi,
  8, 86

\bibitem[{Laginja \& Wakeford(2020{\natexlab{a}})}]{Laginja2020}
Laginja, I., \& Wakeford, H.~R. 2020{\natexlab{a}}, Journal of Open Source
  Software, 5, 2281, \dodoi{10.21105/joss.02281}

\bibitem[{Laginja \& Wakeford(2020{\natexlab{b}})}]{iva_laginja_2020_3923986}
---. 2020{\natexlab{b}}, ExoTiC-ISM v2.0.0, v2.0.0,  Zenodo,
  \dodoi{10.5281/zenodo.3923986}

\bibitem[{{Magic} {et~al.}(2015){Magic}, {Chiavassa}, {Collet}, \&
  {Asplund}}]{2015A&A...573A..90M}
{Magic}, Z., {Chiavassa}, A., {Collet}, R., \& {Asplund}, M. 2015, \aap, 573,
  A90, \dodoi{10.1051/0004-6361/201423804}

\bibitem[{{Maxted}(2018)}]{2018A&A...616A..39M}
{Maxted}, P.~F.~L. 2018, \aap, 616, A39, \dodoi{10.1051/0004-6361/201832944}

\bibitem[{{Maxted}(2023)}]{2023MNRAS.519.3723M}
{Maxted}, P. F.~L. 2023, \mnras, 519, 3723, \dodoi{10.1093/mnras/stac3741}

\bibitem[{{Neckel} \& {Labs}(1994)}]{1994SoPh..153...91N}
{Neckel}, H., \& {Labs}, D. 1994, \solphys, 153, 91, \dodoi{10.1007/BF00712494}

\bibitem[{{Viani} {et~al.}(2018){Viani}, {Basu}, {Ong J.}, {Bonaca}, \&
  {Chaplin}}]{2018ApJ...858...28V}
{Viani}, L.~S., {Basu}, S., {Ong J.}, M.~J., {Bonaca}, A., \& {Chaplin}, W.~J.
  2018, \apj, 858, 28, \dodoi{10.3847/1538-4357/aab7eb}

\bibitem[{{Witzke} {et~al.}(2021){Witzke}, {Shapiro}, {Cernetic}, {Tagirov},
  {Kostogryz}, {Anusha}, {Unruh}, {Solanki}, \& {Kurucz}}]{2021A&A...653A..65W}
{Witzke}, V., {Shapiro}, A.~I., {Cernetic}, M., {et~al.} 2021, \aap, 653, A65,
  \dodoi{10.1051/0004-6361/202140275}

\end{thebibliography}
\bibliographystyle{aasjournal}

\end{document}